\pdfoutput=1
\documentclass[%
reprint,
superscriptaddress,
aps,
prl,
]{revtex4-2}
\usepackage{graphicx}
\usepackage{dcolumn}

\usepackage{hyperref}

\usepackage{siunitx}
\begin{document}

\renewcommand{\figurename}{\textbf{Fig.}}
\renewcommand{\thefigure}{\arabic{figure}}
	
\title{Exploitation-exploration transition in the physics of fluid-driven branching}

\author{J. Tauber}
\affiliation{School of Engineering and Applied Sciences, Harvard University, Cambridge, MA 02138, USA}

\author{J. Asnacios}
\affiliation{School of Engineering and Applied Sciences, Harvard University, Cambridge, MA 02138, USA}
\affiliation{Laboratoire de Physique de l'Ecole normale supérieure, ENS, Université PSL, CNRS, Sorbonne Université, Université de Paris, F-75005 Paris, France}%

\author{L. Mahadevan}
\email{lmahadev@g.harvard.edu}
\affiliation{School of Engineering and Applied Sciences, Harvard University, Cambridge, MA 02138, USA}
\affiliation{Department of Organismic and Evolutionary Biology, Harvard University, Cambridge, MA 02138, USA}
\affiliation{Department of Physics, Harvard University, Cambridge, MA 02138, USA}
	
\begin{abstract}
Self-organized branching structures can emerge spontaneously as interfacial instabilities in both simple and complex fluids, driven by the interplay between bulk material rheology, boundary constraints, and interfacial forcing. In our experiments, injecting dye between a source and a sink in a Hele-Shaw cell filled with a yield-stress fluid reveals an abrupt transition in morphologies as a function of injection rate. Slow injection leads to a direct path connecting the source to the sink, while fast injection leads to a rapid branching morphology that eventually converges to the sink. This shift from an exploitative (direct) to an exploratory (branched) strategy resembles search strategies in living systems; however, here it emerges in a simple physical system from a combination of global constraints (fluid conservation) and a switch-like local material response. We show that the amount of fluid needed to achieve breakthrough is minimal at the transition, and that there is a trade-off between speed and accuracy in these arborization patterns. Altogether, our study provides an embodied paradigm for fluidic computation  driven by a combination of local material response (body) and global boundary conditions (environment).
\end{abstract}

\maketitle


Branching morphogenesis in passive and active matter arises from various driving mechanisms and occurs across a wide range of scales. Examples include viscous fingering in 2-fluid systems~\cite{Saffman1958}, erosion in frangible media~\cite{Cerasi1998,Mahadevan2012}, and hydraulic fracture in yield-stress fluids~\cite{Ball2021} for physical systems, as well as vascularization and arborization for biological systems~\cite{Fleury2001}. Typically, these branched structures self-organize through a series of local events without an explicit pre-pattern guiding the morphodynamics. Here, we present an experimental platform based on a Hele-Shaw geometry to study the guided emergence of fluid-driven branching via the interplay between local material responses and global gradients induced by the boundary conditions using sources and sinks.


\begin{figure}

		\includegraphics{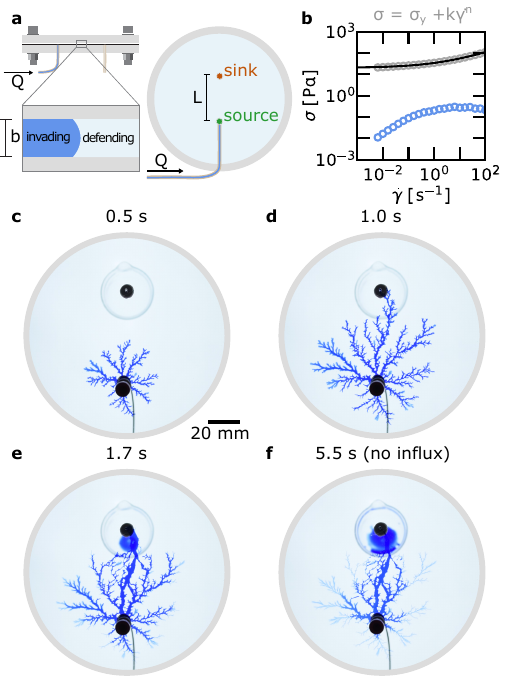} 

        \caption{\textbf{Fluid-driven branching in a Hele-Shaw cell with a point source and a point sink.} \textbf{a}, A sealed Hele-Shaw cell is made of
        acrylic plates separated by a gap $b=\SI{175}{\micro\meter}$. There are two ports, separated by a distance of $L= \SI{60}{\milli\meter}$, which serve as the source and sink. The cell is filled with a yield-stress fluid (gray) and a water-based dye (blue) is injected at a constant flow rate $Q$ from one port (the source). \textbf{b} The shear stress as a function of shear rate for the yield-stress fluid (gray dots), a 2 g/L water-based solution of Carbopol, is well described by the Herschel-Bulkley model (black line) with parameters $\sigma_y =  \SI{20}{\pascal}$, $n = \num{0.40}$, and $k = \SI{13.8}{\pascal\second\tothe{n}}$. In contrast, the dye phase exhibits shear-thinning behavior with significantly lower viscosities (blue line) (see Extended Data Fig.~1). At relatively high injection rates, a pattern develops and progresses through distinct stages: \textbf{c}, branching; \textbf{d}, breakthrough to the sink; \textbf{e}, coalescence of branches that reinforces the source-sink path; \textbf{f}, 
        retraction of other branches---shown here for an injection rate of \SI{16.9}{\milli\liter\per\minute}, corresponding to a shear rate $\dot{\gamma}\approx2Q / (\pi L b^2)\approx \SI{1e2}{\per\second}$.} 
        
        \label{fig:experimentalsetup}
        
\end{figure}

The experiments are conducted in a flow-cell (Fig.~\ref{fig:experimentalsetup}a) consisting of two parallel circular acrylic plates separated by a gap of $b=\SI{175}{\micro\meter}$, filled with a defending fluid. We inject an invading fluid through a point source, and allow the fluids to escape the cell through a point sink placed a distance $L=\SI{60}{\milli\meter}$ away from the source. The flow cell is sealed at the edge; thus, the sink is the only place where fluids can escape (see Methods for a detailed description of flow-cell assembly and operation). This design follows An \textit{et al.}~\cite{An2022} and is a variation of the standard Hele-Shaw cells that are used for investigating viscous fingering at the lab-scale~\cite{Homsy1987}. Unlike these traditional designs that minimize boundary effects by distancing the sink from the source and using symmetrical geometries, we intentionally place the sink and boundaries close to the source and confine the outflow to a single point sink by sealing the boundaries.

\begin{figure*}

		\includegraphics{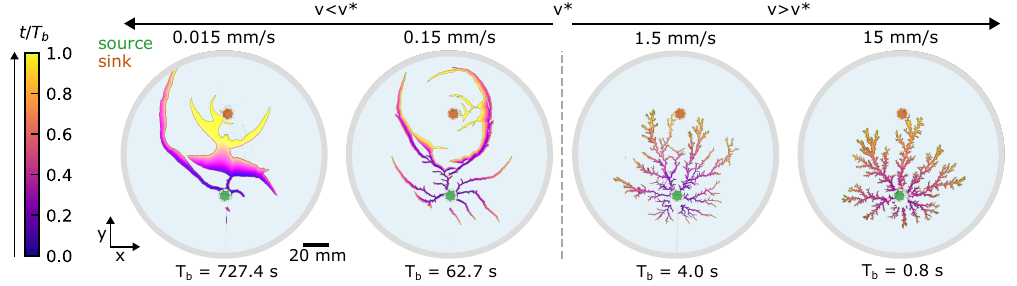}

        \caption{\textbf{Transition from exploitative to exploratory arborization.} Dye is injected from a point source (green) into a Hele-Shaw cell filled with a yield-stress fluid, forming branched patterns that grow until they reach a point sink (orange) located $L=\SI{60}{\milli\meter}$ from the source at the breakthrough time $T_b$. As the invasion speed $v$ increases, a sudden transition in the morphology of the branched structures is observed when $v>v^*$. A color map indicates the temporal evolution of the pattern relative to $T_b$.}
        
        \label{fig:morphology_change}
        
\end{figure*}

We inject a water-based blue dye as the invading phase into a defending yield-stress fluid consisting of a water-based microgel suspension (Carbopol EDT NF 2020). This microgel suspension behaves as a linearly elastic solid with a shear modulus of $G = \SI{108}{\pascal}$ below a yield stress of $\sigma_y = \SI{20}{\pascal}$. Above this yield stress, it flows as a shear-thinning fluid (Fig.~\ref{fig:experimentalsetup}b) closely following the Herschel-Bulkley fluid model~\cite{rheo},
\begin{equation}
\left\{\begin{array}{ll}
	\dot \gamma = 0 \;& \mbox{if } \sigma < \sigma_y \\
	\sigma  = \sigma_y + k \dot \gamma ^{n}& \mbox{else}
\end{array} \right.
\label{eq:HB}
\end{equation}
with a flow index $n=\num{0.40}$, and a consistency $k=\SI{13.8}{\pascal\second\tothe{n}}$. Note that the invading phase is also shear-thinning, but has a lower viscosity than the defending phase at all shear rates (see Extended Data Fig.~1 for further rheological characterization).

Our experiments show that the interface between the invading and defending phase is unstable (Supplementary Video 1). As the invading phase is injected at the source, branched patterns with sharp tips appear, similar to those observed in hydraulic fracture of microgel suspensions~\cite{Ball2021}, clay suspensions~\cite{vendhame}, and oil-in-water suspensions~\cite{Kawaguchi2004}. Notably, this initial branching behavior persists even in the absence of a sink (see Extended Data Fig.~2 and Supplementary Video 2), suggesting that plate deflection may play a critical role, a point we will come back to. 
Injection past the primary stage of interface destabilization causes branch growth through branch widening and tip splitting (Fig.~\ref{fig:experimentalsetup}c). This process continues until one of the branch tips connects with the sink (Fig.~\ref{fig:experimentalsetup}d). Once a flow path is established between the source and the sink, branching stops, and the pattern reorganizes dramatically by merger of branch tips with nearby branches and withdrawal of fluid from deciduous branches (Fig.~\ref{fig:experimentalsetup}e and Fig.~\ref{fig:experimentalsetup}f).
Reorganization during and after breakthrough is reminiscent of breakthrough-induced loop formation in immiscible viscous fingering~\cite{Zukowski2024} and breakthrough-induced reorganization in miscible viscous fingering within a sealed Hele-Shaw cell~\cite{An2022}. However, a key qualitative difference in our system is that the loops remain stable due to the yield-stress behavior of the defending phase.

Fig.~\ref{fig:morphology_change} and Supplementary Video 1 demonstrate a striking transition in the flow pattern as the injection rate is varied over the range $Q  = \SIrange{30}{30000}{\micro\liter\per\minute}$, or equivalently, the invasion speed $v=Q / (\pi L b)$. To characterize this transition in branching morphodynamics, we define a transitional invasion speed $v^* = \SI{0.5}{\milli\meter\per\second}$  based on a dip in the fractal dimension of the pattern from \num{1.60} to \num{1.35} that occurs at intermediate invasion speeds. This minimum results from an increase in branch width for $v<v^*$ and an increase in the number of tips for $v>v^*$ (see Extended Data Fig.~3 and Supplementary Information Section~1 for further morphological characterization). In Fig.~\ref{fig:morphology_change}, we use a color scale to visualize the morphodynamic process leading up to breakthrough. For $v<v^*$, branches propagate away from the sink, followed by branch widening, resulting in out-flow of the yield-stress phase from the sink. As the pattern approaches the sink, the dominant branch turns around the sink before making contact. For $v>v^*$, the pattern expands more isotropically, with no noticable outflow of the defending phase before breakthrough. Interestingly, the characteristic morphology observed for $v<v^*$ disappears when the sink is removed (see Extended Data Fig.~2).

These observations suggest that the source-sink separation and injection rate conspire to guide branching morphodynamics, with the transversely confining plates determining the transition in breakthrough dynamics at $v^*$. For rigidly confining plates, mass conservation of an incompressible fluid implies that fluid injection will lead to a relatively direct path from source to sink, representing purely exploitative behavior. Conversely, fluid injection in the absence of any confinement will lead to isotropic expansion, representing purely exploratory behavior with locally-guided expansion. Soft confinement by flexible plates interpolates between these extremes;  at low injection rates ($v<v^*$), the plates resist deflection, while at high injection rates ($v>v^*$) the pressure inside the cell is high enough to deflect the plates. This causes a shift from globally-guided (exploitative) to locally-guided (exploratory) behavior as invasion speed increases due to the interplay between bulk material properties that control the local response and boundary conditions that guide the pattern globally.

\begin{figure}

    \includegraphics{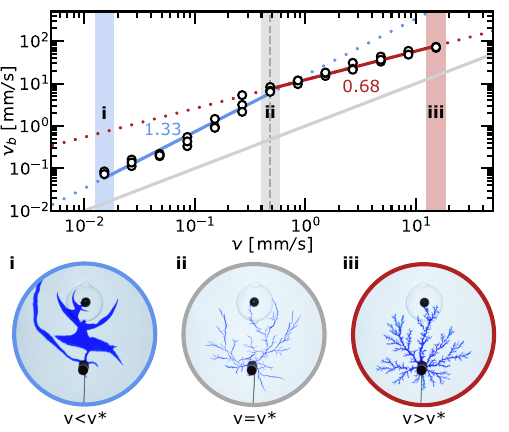}
    
    \caption{\textbf{Breakthrough time scaling reflects arborization transition.} The breakthrough time from the source to the sink, $T_b$, serves as a scalar measure for the transition in branching behavior. To directly compare with the mean invasion speed of the stable front, the experimental breakthrough speed $v_{b} = L / T_b$ is plotted against $v$ for a series of experiments with three repeats. The gray line indicates $v$, the lower bound for the breakthrough speed. We observe a transition in the scaling of $v_{b}$ around the invasion speed $v^* \approx \SI{0.5}{\milli\meter\per\second}$. By fitting the function $v_b = Bv^A$, we find $A = 1.33 \pm 0.06$ and $B = 157 \pm 90 \, \text{m/s}^{1-A}$ for $v <v^*$, while we find $A = 0.68 \pm 0.02$ and $B = 1.31 \pm 0.18 \, \text{m/s}^{1-A}$ for  $v > v^*$. The errors represent the standard deviation in the power law fit, as described in the Methods section. Snapshots of the branched pattern at breakthrough for \textbf{i}, $v=\SI{0.015}{\milli\meter\per\second}$, \textbf{ii}, $v=\SI{0.5}{\milli\meter\per\second}$, and \textbf{iii}, $v=\SI{15}{\milli\meter\per\second}$.}   
    
    \label{fig:time}
    
\end{figure}

To quantify how the interplay between bulk material properties and boundary conditions influences breakthrough dynamics, we plot the breakthrough speed $v_b = L / T_b$, where $T_b$ is the breakthrough time for the fluid to reach the sink, versus the invasion speed $v$ in Fig.~\ref{fig:time}. Here, the invasion speed $v = L / T$, represents the mean speed of a stable unobstructed fluid-fluid interface spreading over a distance $L$ in a time $T$, with $T$ the characteristic time of injection $T = \pi L^2 b / Q$. The experimentally measured breakthrough speeds of unstable branching fronts always exceed this invasion speed as indicated by the gray line in Fig.~\ref{fig:time}. Moreover, we find a transition in the power-law behavior at $v\sim v^*$ with
\begin{equation}
\left\{\begin{array}{ll}
	v_b \sim v^{1.33\pm0.06} \;& \mbox{if } v < v^* \\
    v_b \sim v^{0.68\pm0.02} \;& \mbox{if } v > v^*
\end{array} \right.
\label{eq:scaling_exp}
\end{equation}

The deviation from the scaling $v_b \sim L/ T \sim v$ and the presence of a transition, suggest that a single time scale, here the time scale of injection $T$, is not sufficient to describe the breakthrough process. Three other time scales could potentially affect breakthrough: the Darcy time scale for flow of the invading phase into cracks and fingers $ t_i \sim \frac{\eta_i L^2}{b^2 \Delta P}$, the Darcy time scale for flow of the defending phase towards the sink $t_d \sim \frac{\eta_d L^2}{b^2 \Delta P} \sim \frac{nk \left( \frac{v}{b} \right)^{n-1} L^2}{b^2 \Delta P}$, and the relaxation time of plate deflection coupled to a yield-stress fluid, $t_r \sim \frac{\eta_d}{B} \frac{L_{plate}^6}{b^3} \sim \frac{nk \left( \frac{v}{b} \right)^{n-1}}{B} \frac{L_{plate}^6}{b^3}$ with the invading fluid viscosity $\eta_i$, defending fluid viscosity, $\eta_d$, and flexural rigidity of the flow-cell plates $B \sim E_p t^3$ (see SI Section~3 for details).

To investigate how invasion speed, Darcy flow, and plate deflection interact, we consider a minimal 1D model for invasion and breakthrough of a stable front in a deformable channel (see Supplementary Information Section~4 for details). This model predicts two distinct deflection regimes based on the invasion speed. The transition between these regimes occurs when $t_d = T$, as the deflection $h/b$ reaches the yield strain of the defending phase $\epsilon_y = \sigma_y / G$ (see Extended Data Fig.~5).
The transitional invasion speed follows from the relation $t_d \sim T \sim L/v^*$. At $v^*$, the yield stress and the pressure are linked by a force balance involving the transverse deflection of the plate, expressed as $\Delta P \sim \frac{B}{L_p^4} \frac{\sigma_y b}{G}$. This yields:
\begin{equation}
    v^* \sim \left( \frac{ B}{L_p^4} \frac{b^{n+2}}{ 2^{n-1} n k L} \frac{\sigma_y}{G} \right)^{1/n} \;\;.
    \label{eq:predictionvstar}
\end{equation}
Substituting experimental parameters into this minimal 1D model, we predict $v^*=\SI{1.0}{\milli\meter\per\second}$, qualitatively consistent with our observations of a transitional invasion speed of $\SI{0.5}{\milli\meter\per\second}$ (see Supplementary Information Section~4 for details). 
Numerical solutions of the 1D model across a range of invasion speeds show that, for all $v$, the yield-stress behavior of the defending phase initially prevents outflow until the deflection exceeds the yield strain. When $v<v^*$, any additional influx is compensated by outflow, leading to a deflection that plateaus around $\epsilon_y$. On the other hand, when $v>v^*$, the increased hydraulic resistance amplifies the plate deflection, causing it to surpass $\epsilon_y$.

In our 1D model the transition between deflection regimes has only a minor effect on the breakthrough speed $v_b$, as it considers the propagation of a stable front. In contrast, our quasi-2D experiments exhibit unstable arborization driven by heterogeneities in material properties. 
At high invasion speeds ($v>v^*$), we expect the relaxation dynamics of the fluid-coupled plates to dominate propagation of the branches, so that $v_b \sim L/t_r$. Indeed, we find that $L/t_r \sim v^{1-n} \sim v^{0.6}$ for a flow index of $n=0.40$, consistent with the observed scaling shown in Fig.~\ref{fig:time}. For $v<v^*$, interfacial growth is dominated by the propagation of a single branch tip. Similar to the 1D model, the tip grows in response to outflow of the defending phase. However, in the 2D system the branch can also widen or extend away from the sink, which delays breakthrough. This behavior is captured by the observed scaling $v_b \sim v^{1.33}$, as shown in in Fig.~\ref{fig:time}.

To visualize this role of bulk flow in the arborization transition, we suspended $\SI{50}{\micro\meter}$ particles in Carbopol and performed Particle Image Velocimetry to extract the velocity fields in the defending phase during dye injection. For an invasion speed of $v = \SI{0.015}{\milli\meter\per\second}<v^*$ the flow is localized between the evolving branch pattern and the sink. In contrast, for $v = \SI{15}{\milli\meter\per\second}>v^*$, the flow around the sink is negligible (see Extended Data Fig.~6 and Supplementary Video~3). These observations confirm that the change in branching morphodynamics at the arborization transition is driven by a change in bulk-flow behavior. For $v<v^*$ the flow field between the source and the sink guides branch propagation, directing the invading phase towards the breakthrough path. However, when $v>v^*$ the absence of localized flow in the defending phase effectively screens the sink from the advancing branch tips, resulting in a more homogeneous distribution of the invading phase across the interface.

Building on these findings, we estimate the volume of fluid contributing to the breakthrough path $V_{b,g}$ and the effective flow-rate $Q_g$ driving its formation. From this, we can estimate the breakthrough speed as $v_b = \frac{L Q_{g}}{V_{b,g}}$ (see Extended Data Fig.~7 and Supplementary Information Section~1 for details). This approach enables us to use snapshots at breakthrough to deduce the scaling relations:
\begin{equation}
\left\{\begin{array}{ll}
	v_b \sim v^{1.40\pm0.07} \;& \mbox{if } v < v^* \\
    v_b \sim v^{0.73\pm0.05} \;& \mbox{if } v > v^*
\end{array} \right.
\end{equation}
These results are consistent with the direct measurements of the breakthrough time given by Eq.~(\ref{eq:scaling_exp}), and confirm that the transition in breakthrough dynamics is a consequence of the arborization transition.

\begin{figure*}

    \includegraphics{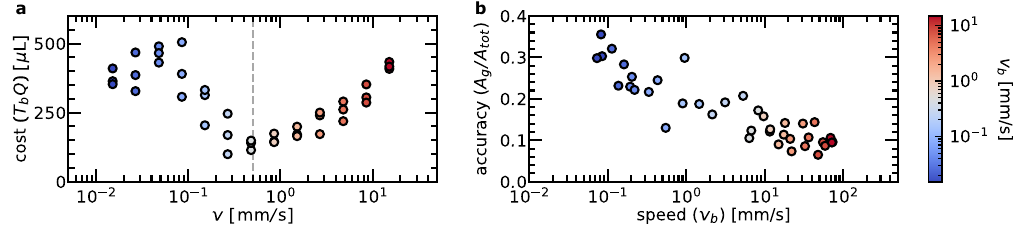}
    
    \caption{\textbf{Characterizing optimal breakthrough.} \textbf{a}, Total volume injected up to breakthrough, $T_b Q$, showing a minimum around $v = v^*$ (dashed gray line). This indicates that in a process where material cost is a concern, the optimal operating point occurs at $v = v^*$. \textbf{b}, Optimal performance represents a trade-off between cost, speed, and accuracy. While the cost, measured as injected volume, is minimized at intermediate invasion speeds, the accuracy—defined as the ratio of the area covered by the geodesic path to the total branch area—decreases with increasing breakthrough speed. Marker color indicates the breakthrough speed, as shown by the color scale.}
    
    \label{fig:optimalbreakthrough}
    
\end{figure*}

If we consider fluid-driven arborization as a search for optimal paths connecting the source and the sink, the process can be viewed as a form of physical computation. The observation that only a fraction of the injected volume contributes to breakthrough indicates that invasion speed influences the efficiency of this search process. To quantify this, we assess the solution quality in terms of cost, speed, and accuracy. Specifically, we define the total injected volume at breakthrough, $QT_b$, as the cost (Fig.~\ref{fig:optimalbreakthrough}a). We find that the invasion speed at the arborization transition $v^*$ minimizes the search cost, optimizing spatial and temporal processes. As the invasion speed increases, more branch tips are activated, exploring a larger area of the cell but reducing spatial efficiency. Conversely, at low invasion speeds, the active branch tips take longer to reach the sink, delaying breakthrough. For $v \approx v^*$ these competing processes are balanced. Additionally, defining the accuracy as the ratio of the area covered by the geodesic path connecting the source and the sink, $A_{g}$, to the total area, $A_{tot}$, we observe that accuracy decreases monotonically with increasing speed (Fig.~\ref{fig:optimalbreakthrough}b), resulting in a Pareto-like trade-off.

Our investigation of the self-organized patterns resulting from fluid injection between a point source and a point sink reveals an abrupt transition in arborization morphology, driven by the interplay between local material responses and global boundary-induced gradients. The scaled invasion speed $v/v^*$ determines which of two distinct search strategies emerges: source-driven unguided exploration ($v>v^*$) or source-sink-guided exploitation ($v<v^*$). This switch is likely to be of relevance in flow-driven applications, such as enhanced oil recovery~\cite{Maheshwari2016} and enhanced geothermal systems~\cite{Olasolo2016}. This behavior in a purely physical setting is akin to the exploration-exploitation transition in active systems, such as animal foraging~\cite{Eliassen2007,BergerTal2014} and robotic search problems~\cite{Biswas2023,Munir2023}, where a sentient agent dynamically changes strategy based on sensing and action, using local and global information. In our material analog, information is encoded (in the body) through local stress at the branch tip and global pressure gradients. The injection rate at the boundary (the interface with the environment) controls the response of the system, triggering a switch from (slow) exploitation to (fast) exploration when the rate of (global) information available at the branch tip (analogous to the decision-maker) falls below a threshold.  

It may be useful to close by contextualizing our study as an example of physics as an embodied computational engine. We begin by noting that the workhorse of neural networks, the rectified linear unit (ReLU)~\cite{Householder1941} arises naturally in yield-stress fluids, which exhibit a similar local response. When combined with the global response determined by boundary conditions and system geometry, our system might be thought of as a continuum analog of a neural network, an example of physical reservoir computing~\cite{Nakajima2020}, or a physical instantiation of how aneural organisms, such as slime-mold, exploit physics to solve search problems~\cite{Tero2007,Reid2012,Murugan2021}. But how we might further generalize and quantify this analogy beyond search and navigation problems remains a question for the future.

\bibliography{references}

\section*{Acknowledgements}
We acknowledge support from the EMBO postdoctoral fellowship ALTF 2022-7, the Harvard NSF DMR 2011754, NSF DMR 1922321, the Simons Foundation, and the Henri Seydoux Foundation for partial financial support. We thank the Neurotechnology Core for the use of their machine shop. Rheology measurements were performed at the Weitzlab Rheology Center supported by the Harvard MRSEC DMR 1420570.

\end{document}